# Magnetism and Superconductivity in $RuM_{1.5}Ce_{0.5}Sr_2Cu_2O_{10}$ (M=Eu and Y)

I. Felner[a], E. Galstyan[a], V.P.S. Awana[b] and E. Takayama-Muromachi[c]

[a]Racah Institute of Physics, The Hebrew University, Jerusalem, 91904, Israel.
[b]NPL, K.S. Krishnan Marg, New Delhi 110012 India
[c]Superconducting Materials Center, NIMS, 1-1 Namiki, Tsukuba, Ibaraki, 305 0044, Japan

$RuY_{1.5}Ce_{0.5}Sr_2Cu_2O_{10}$, is not superconducting (SC) and orders magnetically at $T_M$ =152 K. The Ru moment at 5K is 0.75$\mu_B$. Ferromagnetic hysteresis loops are observed below $T_{irr}$= 100 K. At 100 K, the remanent magnetization ($M_{rem}$) and the coercive field ($H_C$) become zero. Surprisingly, at $T_{irr} <T< T_M$ a reappearance of antiferromagnetic (AFM) hysteresis loops are observed with a peak at 120 K for both $M_{rem}$ and $H_C$. The paramagnetic constants are: $P_{eff}$ =2.06 $\mu_B$ (greater than the 1.73$\mu_B$ expected for low spin $Ru^{5+}$) and $\theta$= 136 K which agrees well with $T_M$. Substitution of Zn (2.5% atm.) for Cu in $RuEu_{1.5}Ce_{0.5}Sr_2Cu_2O_{10}$ ($T_C$=38 K), suppresses $T_C$ but does not affect the magnetic features such as: $T_M$ and $T_{irr}$, $M_{rem}$ and $H_C$. This indicates that both SC and magnetic states are practically decoupled.

We argue that the magnetic structure of this system becomes (i) AFM ordered at $T_M$. (ii) Due to the tilting of the $RuO_6$ octahedra away from the crystallographic c axis, at $T_{irr} < T_M$, weak-ferromagnetic (W-FM) order is induced by the canting of the Ru moments. (iii) At $T_C$ ($T_M/T_C$~4), the specific materials (M=Eu, Gd) become SC. Below $T_C$, both SC and W-FM states coexist.

**Introduction**

The $RuM_{2-x}Ce_xSr_2Cu2O10$ (M=Eu and Gd, Ru-1222) materials display a magnetic transition at $T_M$= 125-180 K and bulk SC below $T_C$ = 32-50 K ($T_M >T_c$) depending on the R/Ce ratio and on the oxygen concentration [1]. X-ray-absorption spectroscopy taken at the K edge of Ru, reveals that the Ru ions are basically $Ru^{5+}$ [2] or $Ru^{4.75+}$ [3]. The SC state in Ru-1222, is well established and understandable. The temperature dependence of the magneto-resistance data yield $H_{c2}(0)$= 39 T and a coherence length of $\zeta(0)$ =140 Å along the $CuO_2$ planes [4]. Due to the granular nature of the materials the critical current density at 5 K is extremely small: $J_C(0)$ =22 A/cm$^2$. [5]. STM studies have demonstrated that all materials are microscopically uniform with no evidence for spatial phase separation of SC and magnetic regions [6].

The published data up to now have not included any determination of the magnetic structure in Ru-1222. The accumulated results are compatible with two alternative scenarios. (A) Going from high to low temperatures, the magnetic behavior is basically divided into two regions. (i)

At $T_M$(Ru), *all* the material becomes AFM ordered. (ii) At $T_{irr}$ (<$T_M$), a W-FM state is induced, which originates from canting of the Ru moments. (iii) At $T_C$ < $T_{irr}$, SC is induced and both the SC and the WFM states coexist intrinsically on a microscopic scale. (B) Detailed analysis of the magnetization under various thermal-magnetic conditions suggests a *magnetic* phase separation into FM and AFM nano-domain species inside the crystal grains. About 10% of the material becomes FM at $T_M$ and persists down to low temperatures, whereas the rest orders AFM at a lower temperature (at the so-called $T_{irr}$) and becomes SC at $T_C$. In this scenario, the unusual SC state is well understood [7].

The aim of this paper is the elucidation of the magnetic properties of the Ru-1222 and it is divided into two parts. (i) We report on Zn substitution for Cu in RuEu$_{1.5}$Ce$_{0.5}$Sr$_2$(Cu$_{1-x}$Zn$_x$)$_2$O$_{10}$ (x=0, 0.01 and 0.025). It is shown that $T_C$ is reduced from 38 K (for x=0), to 26 K for 0.01. The x=0.025 material is not SC down to 4.2 K. On the other hand, the Ru sublattice magnetic state is not affected by the presence or absence the SC state, indicating that the two states are practically decoupled. (ii) We present a detailed magnetization study of none SC RuY$_{1.5}$Ce$_{0.5}$Sr$_2$Cu$_2$O$_{10}$ (Ru-1222Y), which has been synthesized under 6 GPa at 1200°C for 2 hours [8]. The non-magnetic Y ions permit a direct interpretation of the intrinsic Ru magnetism. For all samples studied, wide ferromagnetic hysteresis loops exist at low temperatures. They close themselves around $T_{irr}$, thus $M_{rem}$ and $H_C$ become zero. Surprisingly, at $T_{irr}$<T< $T_M$ a reappearance of the $M_{rem}$ and $H_C$ is observed. This finding can be interpreted only by assuming model A.

**Experimental details**

The samples were prepared by solid-state reactions as described elsewhere [1,8]. Powder X-ray diffraction measurements confirmed the tetragonal structure (SG I4/mmm) and yield the lattice parameters: a=3.819(1) Å and c=28.445 Å for Ru-1222Y and a=3.846(1) Å and c=28.72(1) Å for all Zn doped materials. Dc magnetic measurements were performed in a commercial SQUID magnetometer The normalized real ac susceptibility was measured at ($H_{dc}$=0) by a home-made probe inserted in the SQUID, with an excitation frequency of 733 Hz and an amplitude of 30 mOe.

**Experimental results**

(i) The temperature dependence of the ac susceptibility curves of RuEu$_{1.5}$Ce$_{0.5}$Sr$_2$(Cu$_{1-x}$Zn$_x$)$_2$O$_{10}$ are presented in Fig. 1. It is readily observed that $T_C$ is reduced from 38 K for x=0, to 26 K for 1% at. Zn. For x=0.025, the sample is not SC and is magnetically ordered only. Similar values were obtained by dc magnetic and resistivity measurements. It is reminiscent of the typical behavior of all Zn doped HTSC materials. Since Cu and Zn are both divalent ions, we may assume that all these three materials (which were prepared simultaneously under the same conditions) have the same oxygen concentration. The presence of Zn in the Cu-0 layers, slightly affects the peak position at 80 K.

The isothermal M(H) curves performed at various temperatures, are strongly dependent on the field (up to 2-4 kOe), until a common slope is reached [1]. M(H) can be described as:

(1) $M(H) = M_{sat} + \chi H$,

where the saturation moment ($M_{sat}$) corresponds to the W-FM contribution of the Ru sub lattice, and $\chi H$ is the linear paramagnetic contribution of Eu and Cu. Fig. 2 shows the temperature dependence of $M_{sat}$ for the three materials studied, which becomes zero at $T_M(Ru)=165(2)$, regardless of their Zn concentration. Note the FM-like shape of the curves. The relatively wide FM hysteresis loops obtained at low temperatures ($H_C$ ~400 Oe at 5 K), become narrow as the temperature increases and practically disappear at $T_{irr}$ (80 K). Above $T_{irr}$, small hysteresis loops are reopened, with a maximum of $H_C$ (130 Oe.) and $M_{rem}$ at 120 K for all three samples (Fig. 3).

(ii) The temperature dependence of the ac susceptibility for the non-SC $RuY_{1.5}Ce_{0.5}Sr_2Cu_2O_{10}$, is presented in Fig. 4. Two peaks at 105(1) and 127 (1) K are observed. Similar peaks were observed in the ZFC and FC dc magnetic measurements measured at 15 Oe. $T_M(Ru)=152(2)$ K was obtained from the temperature dependence of the $M_{sat}$.

All isothermal M(H) curves below $T_M$ depend strongly on H. The remarkable feature is the apparent tendency toward saturation but not reaching full saturation even at 5 K and 5 T. This behavior is reminiscent of the unsaturated M(H) curves observed for itinerant ferromagnetic $SrRuO_3$ single crystals at various orientations [9]. The measured moment for Ru-1222Y at 5 K and 5 T, is $M=0.71\mu_B/Ru$. We estimate the Ru moment at infinite H, by plotting $M^2 \alpha 1/H$ (for high H values). Extrapolating to $1/H=0$, yields: $M_{max}= 0.75\mu_B/Ru$, a value which is smaller than $1\mu_B$, the expected moment for $Ru^{5+}$ in the low-spin state (g=2 and S=0.5). Using relation (1) we obtained at 5 K $M_{sat} = 0.63(1)\mu_B$ and $\chi = 1.53*10^{-2}\mu_B/T$. This $\chi$ value is much larger than the Cu ion contribution to the susceptibility.

At low temperatures, at low applied fields the M(H) curves exhibit a typical ferromagnetic-like hysteresis loop (Fig. 5 inset) and values deduced at 5 K are: $M_{rem}= 0.31\mu_B/Ru$ and $H_C = 410$ Oe. Here again, both $M_{rem}(T)$ and $H_C(T)$ decrease with temperature and become zero around $T_{irr} =100$ K (Fig. 5). At higher temperatures reappearance of the hysteresis loops is observed with a peak of $M_{rem}(T)$ and $H_C(T)$ around 125 K (Fig. 6) close to the minor peak in

Fig. 4. In contrast to the FM-like hystersis loop obtained at T< $T_{irr}$, the loops above $T_{irr}$ exhibit an AFM like feature [8].

In the paramagnetic range, the $\chi(T)$ curve (measured at 2T up to 400 K), can be fitted by the Curie-Weiss (CW) law: $\chi =\chi_0 +C/(T-\theta)$, where $\chi_0$ is the temperature independent part of $\chi$, C is the Curie constant, and $\theta$ is the CW temperature. [In order to get the net Ru contribution, we subtracted the $\chi(T)$ of $YBa_2Cu_2O_7$ (measured at 2T) which is roughly temperature independent (1.8-2*10$^{-4}$ emu/mol Oe)]. The values obtained are: $\chi_0$ =0.0014 and C=0.523(1) emu/mol Oe, which corresponds to $P_{eff}$ =2.05$\mu_B$/Ru (1.73$\mu_B$ is the expected value for low-spin $Ru^{5+}$), and $\theta$= 136(1) K which is in fair agreement with $T_M$(Ru).

## Discussion

Figs 2-3 show that the SC materials and the non-SC one (x=0.025) have basically the same $T_M$, $H_C$ and $M_{sat}$. This means that the two SC and magnetic states are decoupled from each other. Without a detailed magnetic structure from neutron diffraction studies, it is difficult to comment on the exact nature of the Ru-1222 system. However, our general picture is that all materials have a similar magnetic structure. The results shown here are compatible with our qualitative model (A), as follows. (i) In the PM range the extracted net $P_{eff}$ =2.05 $\mu_B$/Ru is somewhat greater than the calculated value of low-spin $Ru^{5+}$ (1.73 $\mu_B$) suggesting that the assumption of completely localized moments is not adequate for this system. (ii) At $T_M$ (152-165 K), the Ru sub-lattice becomes AFM ordered at low applied fields. This interpretation is supported by the small peak observed in the ac susceptibility (Fig. 4). By applying a magnetic field, the AFM Ru moments are realigned through a spin-flip process by H, to form the AFM-like shape hysteresis loops as observed in Figs 5. (iii) At $T_{irr}$ (80-100 K), a W-FM is induced, which originates from canting of the Ru moments. This canting arises from the Dzyaloshinsky-Moriya anti-symmetric super-exchange interaction, which by symmetry, follows from the fact that the $RuO_6$ octahedra tilt away from the crystallographic c axis [10]. The ratio of $M_{rem}$ /$M_{max}$ (0.31/ 0.75=0.41) for Ru-1222Y, is also consistent with W-FM order in this system. $M_{max}$ =0.75$\mu_B$/Ru obtained at 5 K is a large fraction of 1 $\mu_B$, implying a very large canting angle (49°) of the AFM Ru moments. Alternatively, the unsaturated magnetization curves at low temperatures may suggest (similar to $SrRuO_3$) itinerant-electron magnetism in Ru-1222. Supporting evidence to this scenario are: (a) the high $P_{eff}/M_{sat}$~3 ratio and (b) the $P_{eff}$ =2.05 $\mu_B$ value which exceeds 1.73 $\mu_B$ expected for a localized $Ru^{5+}$ low spin-state. (iv) For Eu and Gd, SC, which is confined to the $CuO_2$ layers, is induced at $T_C$. Below $T_C$, both SC and W-FM states coexist and the two states are practically decoupled.

Model (B) described above [7], cannot be reconciled with the data presented here: (a) The high 0.75 $\mu_B$/Ru moment at 5 K (b) The continuous $M_{sat}$ curve (Fig. 2) which does not show any inflection at $T_{irr}$ or at lower temperatures (c) The reopening of hysteresis loops above $T_{irr}$ 80-100 K, and as a result the increase of both $M_{rem}$ and $H_C$ (Figs. 3,6). According to model B, the hysteresis loops opened at $T_M$ would remain all the way down to low temperatures and both $M_{rem}$ and $H_C$ would increase continuously, or at least remain constant.

**Acknowledgments** This research was supported by the Israel Academy of Science and Technology and by the Klachky Foundation for Superconductivity.

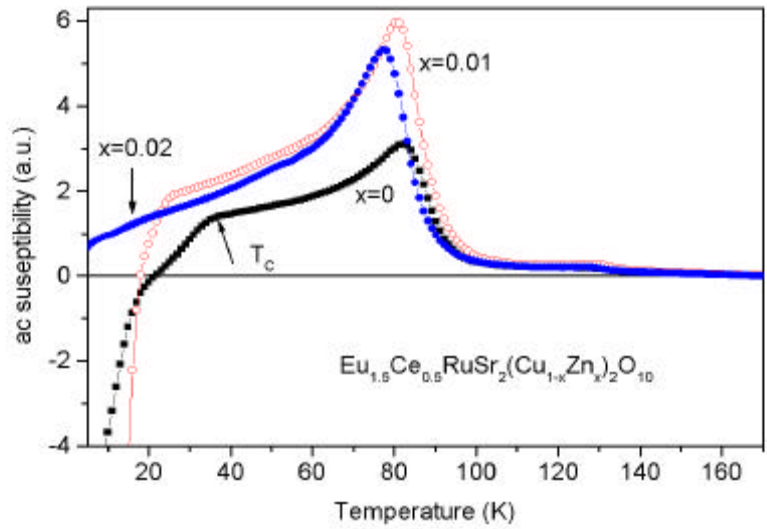

Fig. 1 The temperature dependence of the normalized real ac susceptibility curves of $RuEu_{2-x}Ce_xSr_2(Cu_{1-x}Zn_x)_2O_{10}$

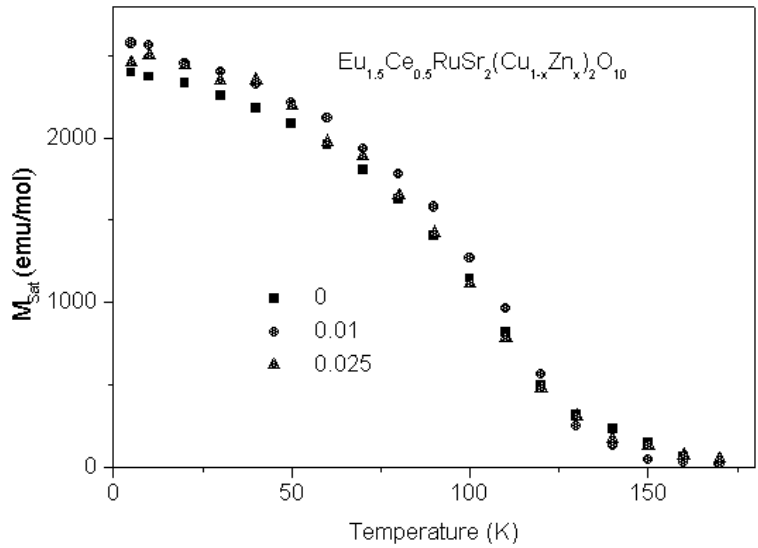

Fig. 2. The temperature dependence of $M_{sat}$ for the Zn doped Ru-1222 materials.

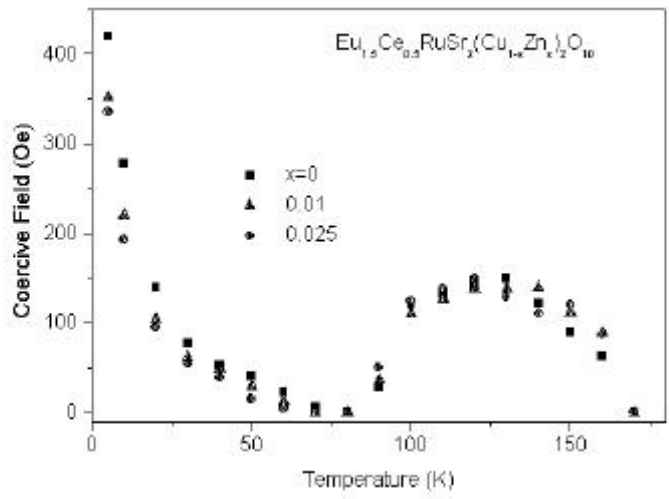

Fig. 3. The temperature dependence of the coercive fields for the three samples

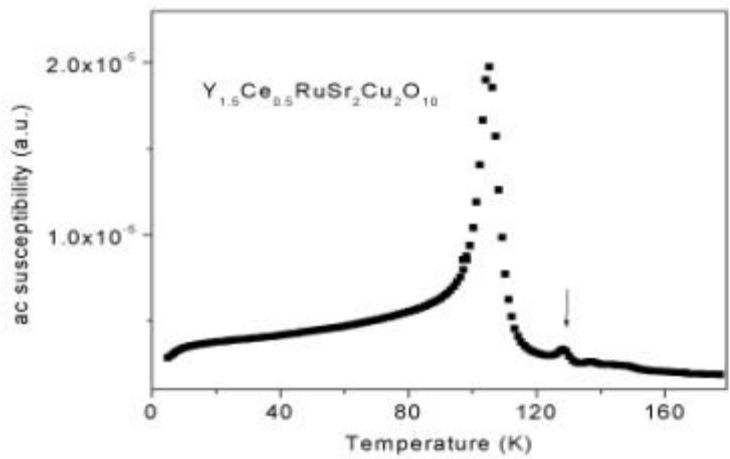

Fig. 4 The temperature dependence of the normalized real ac susceptibility curves of $Y_{1.5}Ce_{0.5}RuSr_2Cu_2O_{10}$

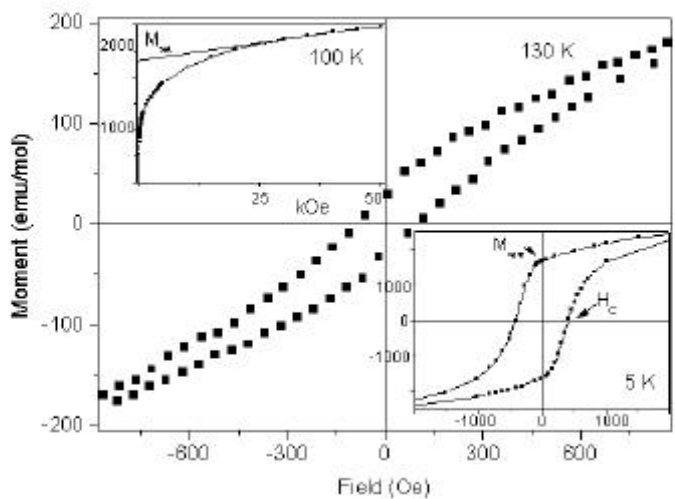

Fig. 5  Hystersis loops of $Y_{1.5}Ce_{0.5}RuSr_2Cu_2O_{10}$ at 5 (inset) and 130 K. Note the absence of hystersis at 100 K (inset).

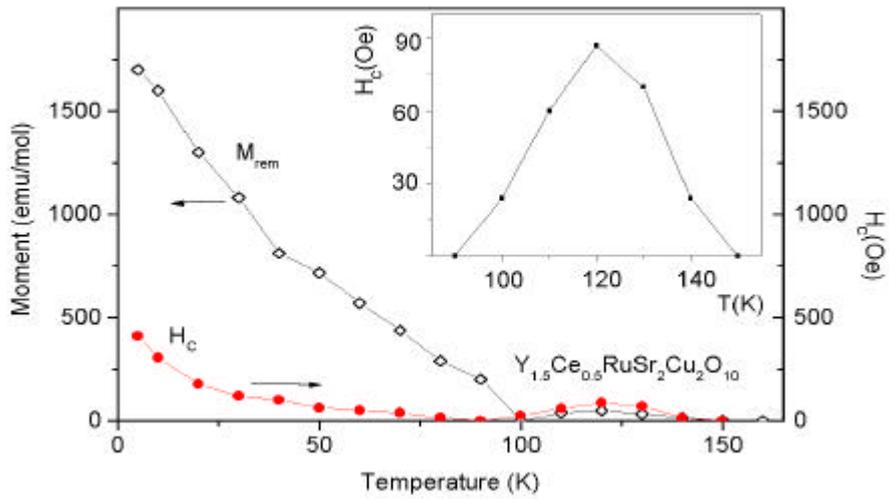

Fig. 6.  The temperature dependence of $M_{rem}$ and $H_C$ for R-1222Y.
The coercive field above $T_{irr}$ is shown in the inset